\documentclass[12pt,a4paper,english]{article}
\usepackage[T1]{fontenc}
\usepackage[latin1]{inputenc}
\usepackage{amsmath}
\usepackage{setspace}
\usepackage{amssymb}

\makeatletter

\providecommand{\LyX}{L\kern-.1667em\lower.25em\hbox{Y}\kern-.125emX\@}

\usepackage{setspace}

\usepackage[numbers]{natbib}

\usepackage[dvips]{graphicx}
\usepackage{epic,epsfig}
\usepackage{subfigure,float}
\usepackage{amsmath,amsfonts,amssymb}
\usepackage{inputenc}
\usepackage{natbib}
\usepackage{euscript}

\usepackage{babel}
\makeatother
\begin{document}
\begin{center}\textbf{\large Dynamic asset trees and portfolio analysis}\end{center}{\large \par}

\begin{center}\textbf{J.-P. Onnela$¹$, A. Chakraborti$¹$, K. Kaski$¹$
and J. Kertész$^{1,2}$}\end{center}

\begin{singlespace}
\begin{center}\textbf{\large $¹$}\textit{Laboratory of Computational
Engineering, Helsinki University of Technology, }\end{center}

\begin{center}\textit{P.O. Box 9203, FIN-02015 HUT, Finland\vskip .4in}\end{center}

\begin{center}\textbf{\large $²$}\emph{Department of Theoretical
Physics, Budapest University of Technology and Economics, }\end{center}

\begin{center}\emph{Budafoki út 8, H-1111, Budapest, Hungary }\end{center}
\end{singlespace}

\noindent \textbf{Abstract}

\noindent The minimum spanning tree, based on the concept of ultrametricity,
is constructed from the correlation matrix of stock returns and provides
a meaningful economic taxonomy of the stock market. In order to study
the dynamics of this asset tree we characterize it by its normalized
length and by the mean occupation layer, as measured from an appropriately
chosen center. We show how the tree evolves over time, and how it
shrinks particularly strongly during a stock market crisis. We then
demonstrate that the assets of the optimal Markowitz portfolio lie
practically at all times on the outskirts of the tree. We also show
that the normalized tree length and the investment diversification
potential are very strongly correlated.

\noindent \vskip .3in


\noindent \vskip .3in


\noindent \newpage

\noindent Portfolio optimization is one of the basic tools of hedging
in a risky and extremely complex financial environment. Many attempts
have been made to solve this central problem starting from the classical
approach of Markowitz \cite{Mark} to more sophisticated treatments
including spin glass type studies \cite{Gall}. In all of these attempts,
correlations between asset prices play a crucial role. A closely related
problem is that of economic taxonomy. In his recent paper \cite{Man1},
Mantegna suggested the study the clustering of companies by using
the correlation matrix of asset returns such that a simple transformation
of the correlations into distances produces a connected graph. In
the graph the nodes are the companies and the `distances' between
them are obtained from the correlation coefficients and the clusters
of companies are identified by means of the minimum spanning tree.
It turned out that in this way the hierarchical structure of the financial
market could be identified in accordance with the results obtained
by an independent clustering method based on Potts super-paramagnetic
transitions \cite{Kull}. In another paper by Bonanno et al. \cite{Gio},
the time evolution of stock indices was studied and significant changes
in the world economy were identified by using appropriate time horizons
and the minimum spanning tree clustering method. The hierarchical
structure explored by the minimum spanning tree also seemed to give
information about the influential power of the companies. The network
of influence was recently investigated by means of a time-dependent
correlation method \cite{Kas}. Some other attempts have been made
to understand the structure of correlation matrices in a highly random
setting using the theory of random matrices \cite{Lal}.

In this paper, we study the minimum spanning tree determined from
correlations between stock returns and call it an `asset tree'. Although
this asset tree can reveal a great deal about the taxonomy of the
market at a given time, it only represents a snapshot of an evolving
complex system. This evolution is a reflection of the changing power
structure in the market and manifests the passing of different products
and product generations, new technologies, management teams, alliances
and partnerships, amongst many other things. This is why exploring
the asset tree \emph{dynamics} can provide us new insights to the
market. Here, by studying the time evolution of the asset tree we
show that although the structure of the tree changes with time, the
companies of the optimal Markowitz portfolio are always on its outer
leaves. We also study the robustness of the tree topology and the
consequences of the market events on its structure. The minimum spanning
tree, as a strongly pruned representative of asset correlations is
found to be robust and descriptive of stock market events.

We start our analysis by assuming that there are $N$ assets with
price $P_{i}(t)$ for asset $i$ at time $t$. Then the logarithmic
return of stock $i$ is $r_{i}(t)=\ln P_{i}(t)-\ln P_{i}(t-1)$, which
for a certain consecutive sequence of trading days forms the return
vector $\boldsymbol r_{i}$. In order to characterize the synchronous
time evolution of stocks, we use the equal time correlation coefficients
between stocks $i$ and $j$ defined as 

\begin{equation}
\rho _{ij}=\frac{\langle \boldsymbol r_{i}\boldsymbol r_{j}\rangle -\langle \boldsymbol r_{i}\rangle \langle \boldsymbol r_{j}\rangle }{\sqrt{[\langle \boldsymbol r_{i}^{2}\rangle -\langle \boldsymbol r_{i}\rangle ^{2}][\langle \boldsymbol r_{j}^{2}\rangle -\langle \boldsymbol r_{j}\rangle ^{2}]}},\end{equation}

\noindent where $\left\langle ...\right\rangle $ indicates a time
average over the trading days included in the return vectors. These
correlation coefficients forming an $N\times N$ matrix with $-1\leq \rho _{ij}\leq 1$,
is then transformed to an $N\times N$ distance matrix with elements
$d_{ij}=\sqrt{2(1-\rho _{ij})}$, such that $2\geq d_{ij}\geq 0$,
respectively. The $d_{ij}$s fulfill the requirements of distances,
even those of ultrametricity \cite{Man1}. We now use the distance
matrix to determine the minimum spanning tree (MST) of the distances,
denoted by $\mathbf{T}$, which is a simply connected graph that connects
all the $N$ nodes of the graph with $N-1$ edges such that the sum
of all edge weights, $\sum _{(i,j)\in \mathbf{T}}d_{ij}$, is minimum.
It should be noted that in constructing the minimum spanning tree,
we are effectively reducing the information space from $N(N-1)/2$
separate correlation coefficients to $N-1$ tree edges.

The dataset we have used in this study consists of daily closure prices
for 116 stocks of the S\&P 500 index \cite{sp}, which were obtained
from the Yahoo website \cite{yahoo}. The time period of this data
extends from the beginning of 1982 to the end of 2000 including a
total of 4787 price quotes per stock, after the removal of a few days
due to incomplete data. We divide this data into $M$ \emph{windows}
$t=1,\, 2,...,\, M$ of width $T$ corresponding to the number of
daily returns included in the window. Different windows overlap with
each other, the extent of which is dictated by the window step length
parameter $\delta T$, describing the displacement between two consecutive
windows, measured also by the number of trading days. The choice of
the window width is a trade-off between too noisy and too smoothed
data for small and large window widths, respectively. In our studies,
$T$ was set to be typically between 500 and 1500 trading days, i.e.,
2 and 6 years, and $\delta T$ to one month including about 21 trading
days. This is in accordance with the suggestions of the Basel committee
\cite{Basel}.

In order to study the temporal state of the market we define the \emph{normalized
tree length} as

\begin{equation}
L(t)=\frac{1}{N-1}\sum _{d_{ij}\in \mathbf{T}^{t}}d_{ij},\end{equation}

\noindent where $t$ denotes the time at which the tree is constructed,
and $N-1$ is the number of edges present in the MST. To characterize
the position of companies in the graph, i.e., the layers on which
the different nodes are located at a given time, we introduce the
concept of a \emph{central node}. Although there is arbitrariness
in the choice of the central node, we propose that it is central in
the sense that any change in its price strongly affects the course
of events in the market on the whole. Thus the central node would
be the company which is most strongly connected to its nearest neighbors
in the tree. With this choice the sum of the correlation coefficients
calculated for the incident edges would be maximum, and/or have the
highest \emph{vertex degree} (the number of edges which are incident
with the vertex). It is also noted that one can have either a static
(fixed at all times) or a dynamic (continuously updated) central node,
without considerable effects on the results. In our studies, General
Electric (GE) was chosen as the central node, since for about 70\%
of the period considered it was the most connected node. A typical
asset tree is shown in Figure 1, where it is evident that companies
become clustered by business sectors.

Figures 2 (a) and (b) show how the normalized tree length $L$ and
the mean correlation coefficient, defined as $\bar{\rho }=\frac{1}{N(N-1)/2}\sum \rho _{ij}$,
where we consider only the non-diagonal and independent $\rho _{ij}$,
evolve with time. The two curves, indeed, look like mirror images,
which is corroborated by the fact that the correlation coefficient
is $-0.96$, indicating that the minimum spanning tree is a strongly
reduced representative of the whole correlation matrix and bears the
essential information about asset correlations. As further evidence
that the MST retains the salient features of the stock market, it
is noted that the 1987 market crash can be quite accurately seen in
Figure 2. The two sides of the ridge actually converge as a result
of extrapolating the window width $T\rightarrow 0$ \cite{jpo}. In
Figure 2 (a), the mean correlation of stocks is very high during the
crash. This is because the market forces act strongly on all the stocks
and force the market to behave in a unified way. Figure 2 (b) also
strengthens this fact: $L(t)$ decreases indicating that the nodes
on the graph are drawn closer together. In order to characterize the
spread of nodes on the graph, we introduce the quantity of \emph{mean
occupation layer} as 

\begin{equation}
l(t)=\frac{1}{N}\sum _{i=1}^{N}\mathop {\mathrm{lev}}(v_{i}^{t}),\end{equation}

\noindent where $\mathop {\mathrm{lev}}(v_{i})$ denotes the level
of vertex $v_{i}$ in relation to the central node, whose level is
taken to be zero. We find that $l(t)$ reaches a very low value at
the time of a market crisis (see Figure 3).

Next, we apply the above discussed concepts and measures to portfolio
analysis. We consider a minimum risk Markowitz portfolio $P(t)$ with
the asset weights $w_{1},\, w_{2},\ldots ,\, w_{N}$. In the Markowitz
portfolio optimization scheme financial assets are characterized by
their average return and risk, both determined from historical price
data, where risk is measured by the standard deviation of returns.
The aim is to optimize the asset weights so that the overall portfolio
risk is minimized for a given portfolio return \cite{soft}. In the
minimum spanning tree framework, the task is to determine how the
assets are located with respect to the central node. Intuitively,
we expect the weights to be distributed on the outskirts of the graph.
In order to describe what happens, we define a single measure, the
\emph{weighted portfolio layer} as

\begin{equation}
l_{P}(t)=\sum _{i\in P}w_{i}\mathop {\mathrm{lev}}(v_{i}^{t}),\end{equation}

\noindent where we have the constraint $w_{i}\geq 0$ for all $i$,
since we assume that there is no short-selling.

Figure 3 shows the behaviour of the mean layer $l(t)$ and the weighted
minimum risk portfolio layer $l_{P}(t)$. We find that the portfolio
layer is higher than the mean layer practically at all times. The
difference in layers depends to a certain extent on the window width:
for $T=500$ it is about $0.76$ and for $T=1000$ about $0.97$.
As the stocks of the minimum risk portfolio are found on the outskirts
of the graph, we expect larger graphs (higher $L$) to have greater
\emph{diversification potential}, i.e., the scope of the stock market
to eliminate specific risk of the minimum risk portfolio. In order
to look at this, we calculated the mean-variance frontiers for the
ensemble of $116$ stocks using $T=500$ as the window width. In Figure
2 (c), we plot the level of portfolio risk as a function of time,
and find a striking similarity between the risk curve and the curves
of the mean correlation coefficient $\bar{\rho }$ and normalized
tree length $L$ of Figures 2 (a) and (b). The correlation between
the risk and $\bar{\rho }$ is $0.82$, while the correlation between
the risk and $L$ is $-0.90$. Therefore, the latter result explains
the diversification potential of the market better.

Finally, in order to investigate the robustness of the minimum spanning
tree topology, we define the survival ratio of tree edges (fraction
of edges is found common in both graphs) at time $t$ as \[
\sigma _{t}=\frac{1}{N-1}|E^{t}\cap E^{t-1}|.\]
 In this $E^{t}$ refers to the set of edges of the graph at time
$t$, $\cap $ is the intersection operator and $|...|$ gives the
number of elements in the set. Under normal circumstances, the graphs
at two consecutive time windows $t$ and $t+1$ (for small values
of $\delta T$) should look very similar. Whereas some of the differences
can reflect real changes in the asset taxonomy, others may simply
be due to noise. We find that as $\delta T\rightarrow 0$, $\sigma _{t}\rightarrow 1$
\cite{jpo}, indicating that the graphs \emph{are} stable in the limit,
and hence our portfolio analysis is justified.

In summary, we have studied the dynamics of asset trees and applied
it to portfolio analysis. We have shown that the tree evolves over
time and have found that the normalized tree length decreases and
remains low during a crash, thus implying the shrinking of the asset
tree particularly strongly during a stock market crisis. We have also
found that the mean occupation layer fluctuates as a function of time,
and experiences a downfall at the time of market crisis due to topological
changes in the asset tree. As for the portfolio analysis, it was found
that the stocks included in the minimum risk portfolio tend to lie
on the outskirts of the asset tree: on average the weighted portfolio
layer is about $1$ level higher, or further away from the central
node, than mean occupation layer for window width of four trading
years. The correlation between the risk and the mean correlation was
found to be quite strong, though not as strong as the correlation
between the risk and the normalized tree length. Thus it can be concluded
that the diversification potential of the market is very closely related
to the behaviour of the normalized tree length.

\vskip .3in

\noindent \textbf{Acknowledgements}

\noindent J.-P. O. is grateful to European Science Foundation for
the grant to visit Hungary, the Budapest University of Technology
and Economics for the warm hospitality and L. Kullmann for stimulating
discussions. This research was partially supported by the Academy
of Finland, Research Centre for Computational Science and Engineering,
project no. 44897 (Finnish Centre of Excellence Programme 2000-2005)
and OTKA (T029985).

\noindent \vskip .3in

\noindent \textbf{Figure Captions}

\noindent \textbf{Fig. 1 :} A typical asset taxonomy (minimum spanning
tree) graph connecting the examined 116 stocks of the S\&P 500 index.
The graph was produced using four-year window width and it is centered
on January 1, 1998. Business sectors are indicated according to Forbes,
\emph{http://www.forbes.com}. In this graph, General Electric (GE)
was used as a a central node and eight layers can be identified. 

\noindent \textbf{Fig. 2 :} Plots of (a) the mean correlation coefficient
$\bar{\rho }$, (b) the normalized tree length $L$ and (c) the risk
of the minimum risk portfolio, as functions of time. The risk is determined
with weight limits of zero lower bound (no short-selling) and unit
upper bound (any asset may constitute the entire portfolio). For all
plots the window width is $T=500$, i.e., two trading years.

\noindent \textbf{Fig. 3 :} Plots of mean occupation layer $l$ and
weighted portfolio layer $l_{P}$ as functions of time. This plot
is based on the window width $T=1000$, i.e., four trading years.


\begin{thebibliography}{10}
\bibitem{soft}Several software packages based on standard procedures are available.
We used \mbox{\textsc{Matlab}} with Financial Toolbox.
\end{thebibliography}
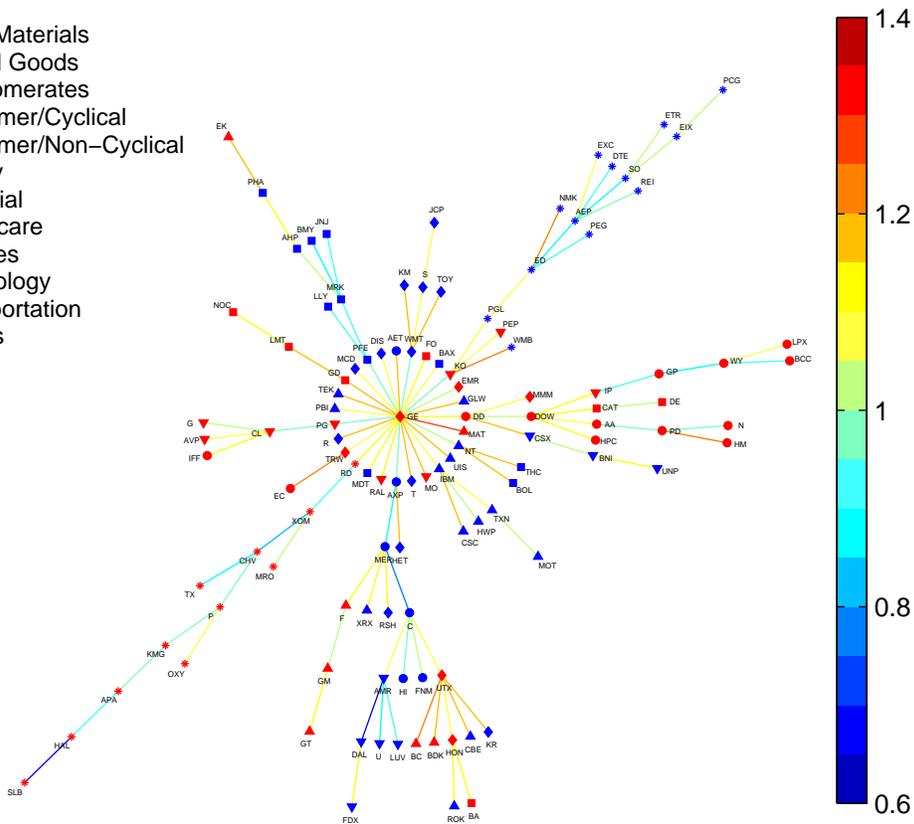
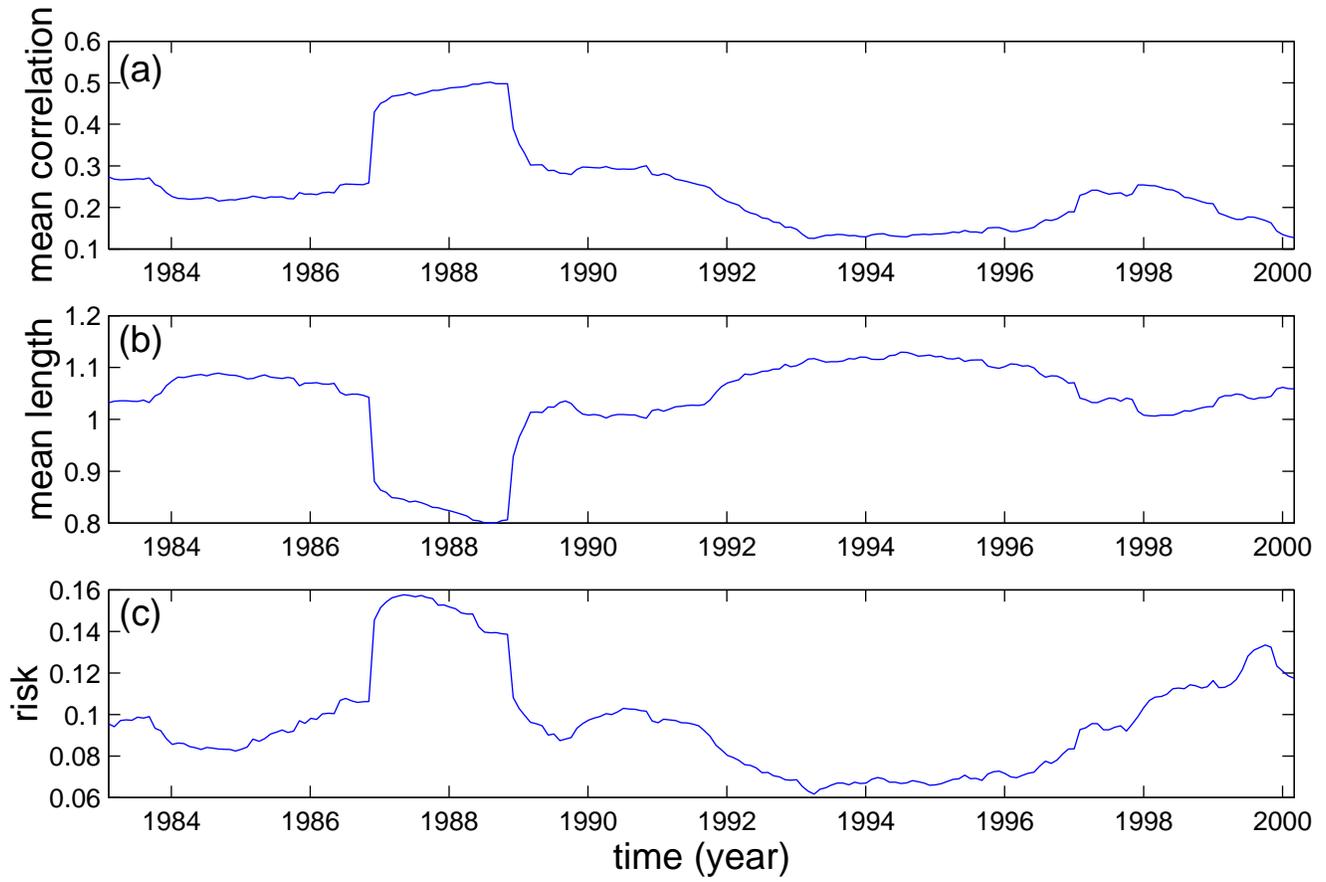
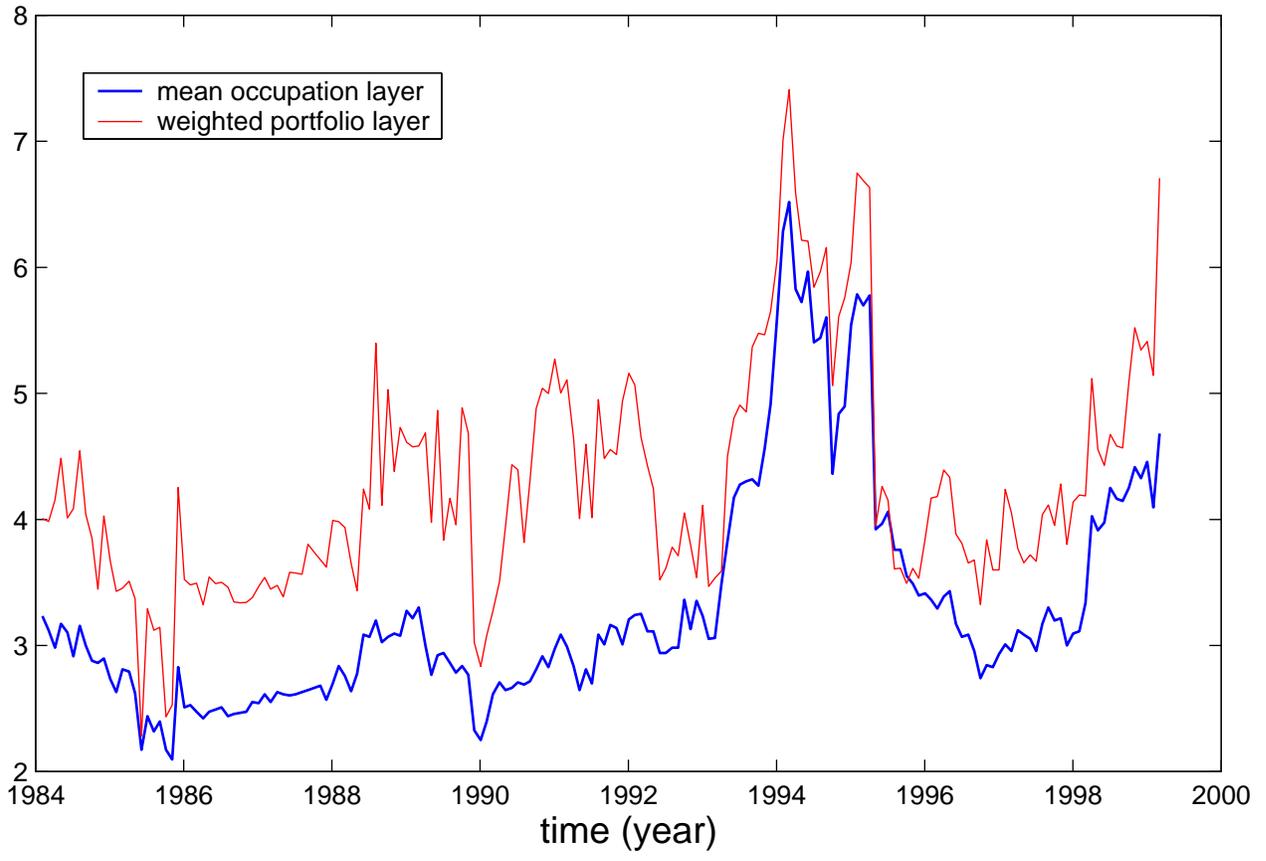
\end{document}